\documentstyle[12pt]{article}
\newcommand{\newc}{\newcommand}
\newc{\gsim}{\lower.7ex\hbox{$\;\stackrel{\textstyle>}{\sim}\;$}}
\newc{\lsim}{\lower.7ex\hbox{$\;\stackrel{\textstyle<}{\sim}\;$}}
\def\NPB#1#2#3{Nucl. Phys. B {\bf#1} (19#2) #3}
\def\PLB#1#2#3{Phys. Lett. B {\bf#1} (19#2) #3}

\def\PRD#1#2#3{Phys. Rev. D {\bf#1} (19#2) #3}
\def\PRL#1#2#3{Phys. Rev. Lett. {\bf#1} (19#2) #3}

\def\ARNP#1#2#3{Ann. Rev. Nucl. Part. Sci. {\bf#1} (19#2) #3}

\begin{document}
\begin{titlepage}
{}.
\vspace{-2cm}
\begin{flushright} RAL-93-010\\UFIFT-93-06\\ March 1993\\
\end{flushright}
\vskip 1.5cm
\begin{center}
{\bf\Large Stitching the Yukawa Quilt}
\vskip 1cm
{\bf P. Ramond}$*$\\
\vskip 2pt
{\it Institute for Fundamental Theory,\\ Department of Physics,
University of Florida,\\
Gainesville Florida 32611 USA}\\
\vskip .25cm
{\bf  R. G. Roberts\\}
\vskip 2pt
{\it Rutherford Appleton Laboratory,\\
Chilton, Didcot, Oxon,
OX11 OQX, UK}\\
\vskip .25cm
{\it and}
\vskip .25cm
{\bf G. G. Ross}\dag\\
\vskip 2pt
{\it Department of Theoretical Physics,\\
Oxford University,\\ Keble Road, Oxford, UK}\\
\end{center}
\vskip .5cm
\begin{abstract}
We develop a systematic analysis of quark mass matrices which,
starting with the measured values of quark masses and mixing
angles, allows for a model independent search for all possible
(symmetric or hermitian) mass matrices having texture zeroes at
the  unification scale.  A survey of all six and five texture
zero structures yields a total of five possible solutions which
may be distinguished by improved measurements of the CKM matrix
elements and which may readily be extended to include lepton
masses with the Georgi-Jarlskog texture. The solutions naturally
suggest a parameterisation for the mass matrices based on a
perturbative expansion and we present some speculations
concerning the origin of such structure.
\end{abstract}
\vskip .5cm
\footnoterule
{\footnotesize $*$ Supported in part by the U.S. Department of
Energy under grant DE-FG05-86ER-40272 \newline \dag SERC Senior
Research Fellow} \end{titlepage}
\setcounter{page}{1}     

\section{\bf Introduction}

The Yukawa sector of the Standard Model is parameterized in terms
of three $3\times 3$ matrices of Yukawa coupling constants. These
serve to determine  the nine quark and lepton masses as well as
the three angles and one phase of the Cabibbo-Kobayashi-Maskawa
(CKM) matrix. Of the masses only one is presently unknown, the
top quark mass although it cannot be heavier than $200$ GeV
without spoiling the consistency of the Standard Model with
experiment.

While the three Yukawa matrices appear as independent parameters
at the level of the Standard Model, they can be correlated in
various theoretical extensions.  In specific GUTs, for example,
there are relations between fermion masses, the best known being
the $SU(5)$ relation $m_b=m_\tau$^\cite{chan} at the unification
scale $M_X$. As with the relation amongst gauge couplings this
relation applies at the unification scale and must be radiatively
corrected. These corrections offer a further tantalising piece
of circumstantial evidence in favour of supersymmetric
unification for, starting with the relation
$m_b(M_X)=m_{\tau}(M_X)$, they can bring the prediction for
$m_b/m_\tau$ into agreement with experiment using the same value
for $M_X$ found in the analysis of gauge couplings^\cite{arse}.
This cannot
be done in the non-supersymmetric case without adding new
interactions.

Further predictions for fermion masses require more sophisticated
GUTs because the simple relations $m_s=m_{\mu}$ and $m_d=m_e$
plus radiative corrections are grossly in disagreement with
experiment. One inspired choice was proposed by Georgi and
Jarlskog^\cite{gj}
and subsequently developed by Harvey, Ramond and
Reiss^\cite{hrr}. Recently it was run
in the SUSY case by Dimopoulos, Hall and Raby^\cite{dhr} and by
Ramond^\cite{pierre} and  Arason, Casta\^no, Ramond and
Piard^\cite{ram}. In this and other^\cite{giud} Ans\"atze the
number of parameters needed to specify the mass matrices is
limited by the requirement that there be ``texture''
zeroes^\cite{wb},\cite{f}. The maximum number of such zeroes in
the up and down quark mass matrices consistent with the absence
of a zero mass eigenvalue is six. With six zeroes there are left
just six real parameters plus one phase to describe the six quark
masses, the three mixing angles and the CP-violating phase and
so one obtains relations between the masses and mixing angles.
It was shown that,
including SUSY radiative corrections, the resulting predictions
for the low energy parameters are in remarkably good agreement
with
experiment.

This illustrates how the analysis of fermion masses can lend
support to the hypothesis of a stage of unification, in the same
way as did the analysis of gauge couplings. In both cases one
uses the renormalisation group to continue the measured values
and looks for simplicity appearing at the unification scale; in
the case of gauge couplings simplicity is equality between the
couplings, in the case of the mass matrices simplicity is simple
ratios of quark and lepton masses and the appearance of
``texture'' zeroes. However the analysis of the mass matrices so
far presented falls short of the ideal ``bottom-up'' approach for
it starts by  assuming a particular, theoretically motivated,
texture for the mass matrices rather than just starting with
measured values, continuing them to high energies, and looking
for simplicity in the form of a definite texture.

In this paper we will attempt to implement this ``bottom-up''
approach through a systematic study of possible zeroes in the
quark Yukawa couplings. The major difficulty in implementing this
program is that laboratory measurements only determine the
masses, i.e. the diagonal mass matrix ${\bf Y}_{u}{Diag},  {\bf
Y}_{d}{Diag}$, and the CKM mass matrix, {\it not} the full mass
matrices, ${\bf Y}_u,{\bf Y}_d$; i.e. we have

\begin{eqnarray}
{\bf Y}_{u}{Diag}={\bf R}_{u}{L}.{\bf Y}_u.{\bf
R}_{u}{R\dagger} \nonumber \\
{\bf Y}_{d}{Diag}={\bf R}_{d}{L}.{\bf Y}_d.{\bf
R}_{d}{R\dagger} \nonumber \\
{\bf V}_{CKM}={\bf R}_{u}{L}.{\bf R}_{d}{L\dagger}
\label{eq:massmat}
\end{eqnarray}

If we are to determine ${\bf R}_{u,d}{L,R}$, and hence the mass
matrices separately, a simplifying assumption is needed.  These
mass matrices come from the Yukawa sector of the theory which,
in the standard model is given by

\begin{equation}
{\cal L}_Y= Q{\bf Y}_u{\bar u}H*+Q{\bf Y}_d{\bar d}H
+L{\bf Y}_e{\bar e}H
\label{eq:ly}
\end{equation}
where $Q_i$ are the three quark doublets, ${\bar u}_i\ ,{\bar
d}_i$ the three right handed charge $-2/3\ ,1/3$ antiquark
isodoublets, $L_i$ the three lepton doublets, ${\bar e}_i$ the
three right handed antilepton isosinglets, and $H$ is the Higgs
doublet normalized to its vacuum value.

Given that the general case is not tractable we will concentrate
in this paper on a promising possibility that has been widely
studied, namely the case that the matrices in family space ${\bf
Y}_u\ ,{\bf Y}_d\ ,{\bf Y}_e$ are symmetric in family space. At
the level of the Standard Model and $SU(5)$, this is not
necessarily true, but at the $SO(10)$ level and beyond, where
each family appears as a single representation, this assumption
is natural (but not inevitable). With this assumption we can
diagonalize the Yukawa matrices, or the associated mass matrices,
by means of a Schur rotation i.e. ${\bf R}L={\bf R}{R
\star}\equiv {\bf R}$.

We can analyze the most general symmetric mass matrix
case\footnote{In fact the analysis applies to hermitian matrices
too for a general $3\times 3$ symmetric mass matrix may be
transformed to an hermitian matrix through the freedom to
redefine the nine phases of the three left-handed doublets and
six right-handed singlets of quark fields. These 9 phases may be
used to make both the up and down quark mass matrices hermitian
since it is always possible to choose a basis in which either the
top or the bottom mass matrix is diagonal.} leading to five or
six texture zeroes, because there are just 6 possible forms of
symmetric mass matrix with an hierarchy of three non-zero
eigenvalues and three texture zeroes (at least one of the up or
down quark mass matrices must have three of the texture zeroes).
Allowing for the redefinition of the quark fields to absorb
phases, these matrices involve just three real parameters. This
allows us to determine, up to the six fold discrete ambiguity,
the diagonalising matrix ${\bf R}_u$ (or ${\bf R}_d$) in terms
of the masses. Hence, using eq(\ref{eq:massmat}), we may compute
${\bf R}_d$ (or ${\bf R}_u$) in terms of the CKM matrix and hence
find ${\bf Y}_d$ (or ${\bf Y}_u$). Further texture
zeroes in ${\bf Y}_u$, ${\bf Y}_d$ will result in
predictions for the mixing angles of the
CKM matrix.

The advantage of this technique is that it allows a determination
of the down quark mass matrix using experimentally measured
quantities without prejudicing the result by the choice of a
specific texture.  Thus the general problem of searching for
structure in mass matrices may be solved with the assumption of
symmetric mass matrices for the case that there are 5 or 6
texture zeroes. We will also consider the remaining case of just
4 texture zeroes, although in this instance there are fewer
predictions making it somewhat uninteresting given the success
of the more predictive forms.

The paper is organised as follows. In Section^\ref{sec:B} we
introduce the procedure needed for a general analysis including
the introduction of a Wolfenstein-like parameterisation for the
mass matrices and the inclusion of the radiative corrections
needed to continue the mass matrices to high energy.
Section^\ref{sec:tex} gives an explicit example of the analysis
and presents the textures consistent with present measurements
of masses and mixing angles. Section^\ref{sec:D} presents the
results of our general analysis in which the radiative
corrections are
determined numerically by integrating the renormalisation group
equations. This allows us to include the effects of thresholds
correctly and to perform a complete analysis of the gauge
couplings, radiative electroweak breaking, and masses and mixing
angles. Section^\ref{sec:E}  discusses these results in the
context of the analytic solutions. Finally
Section^\ref{sec:F} presents our conclusions.

\section{A general analysis for symmetric mass matrix texture.}
\label{sec:B}

The procedure we adopt is straightforward:

\begin{itemize}

\item We first assume that the up quark mass matrix ${\bf Y}_u$
has a certain texture ({\it i.e.} zeroes in specific places)).
After using the freedom to redefine quark phases to make the
elements of ${\bf Y}_u$ real the diagonalising matrix ${\bf R}_u$
is parametrized by three angles\footnote{With this phase
convention it is clear the CP violating phase resides in ${\bf
R}_d$}. If ${\bf Y}_u$ has three zeroes, these angles will be
related to quark mass ratios. If ${\bf Y}_u$ has just two zeroes
there will be one undetermined angle.

\item We now form the down quark mixing matrix by computing
\begin{equation}
{\bf R}_{d}=P_d{\bf V}_{CKM}{\dagger}P_u{\dagger}.{\bf R}_{u}
\label{eq:vd}
\end{equation}
where $P_{u,d}$ are diagonal matrices of phases $e{i \;
\phii_{u,d}}$ needed to express the ``measured'' CKM matrix in
a basis in which the quark fields have arbitrary phases. Using
${\bf R}_d$ we may form the down quark mass matrix
\begin{equation}
{\bf Y}_d={\bf R}_{u}{\dagger}.P_u.{\bf V}_{CKM}.P_d{\dagger}
{\bf Y}_{d}{Diag}.P_d. {\bf
V}_{CKM}{\dagger}.P_u{\dagger}.{\bf R}_{u}
\label{md}
\end{equation}

In writing this we have assumed ${\bf Y}_d$ is hermitian rather
than symmetric and is diagonalised by an hermitian matrix. As
discussed above we are free to do this because we may use the
freedom to redefine quark phases to change the symmetric matrix
to an hermitian matrix.
\item We examine the matrix elements of ${\bf Y}_d$ and derive
relations between quark masses and the CKM mixing angles by
requiring that some of these elements be zero (texture zeroes).

We then start the process all over again, ${\it i.e.}$ input a
different texture for ${\bf Y}_u$, etc.

\item Finally we repeat the whole process starting with the down
quark mass matrix ${\bf Y}_d$ and computing the up quark mass
matrix ${\bf Y}_u$. These steps leave us with all possible
relations among quark masses and mixing angles, derived from
requiring zeroes in the Yukawa matrices.

\item Before comparing with experiment, we run each of these
relations through the renormalization group machine to include
the radiative corrections.
\end{itemize}

This outlines the scheme of analysis we adopt. Its implementation
requires the determination of the possible texture structures,
the parameterisation of the mass matrices in a manner that allows
for a systematic analysis of the predictions, and a determination
of the radiative corrections. We turn now to a discussion of each
of these points.

\subsection{Possible texture structures}
As we mentioned above the analysis of the general case is
possible because there are just 6 possible forms of symmetric
mass matrices with just three non-zero eigenvalues and the
maximum number (three) of texture zeroes capable of describing
the hierarchy of up or down quark mass matrices. These are
\begin{itemize}
\begin{enumerate}
\item
\begin{eqnarray}
\left (
\begin{array}{clcr}
a_1 &0&0\\
0&b_1&0\\
0&0&c_1
\end{array}
\right )
\label{eq:f1}
\end{eqnarray}

\item\footnote{This is the form used in refs
^\cite{gj,hrr,dhr,ram,hh}} \begin{eqnarray}
\left (
\begin{array}{clcr}
0 &a_2&0\\
a_2&b_2&0\\
0&0&c_2
\end{array}
\right )
\label{eq:f2}
\end{eqnarray}

\item
\begin{eqnarray}
\left (
\begin{array}{clcr}
a_3 &0&0\\
0&0&b_3\\
0&b_3&c_3
\end{array}
\right )
\label{eq:f3}
\end{eqnarray}

\item\footnote{This is the form used for up quarks in
ref.^\cite{giud}.} \begin{eqnarray}
\left (
\begin{array}{clcr}
0 &0&a_4\\
0&b_4&0\\
a_4&0&c_4
\end{array}
\right )
\label{eq:f4}
\end{eqnarray}

\item\footnote{This is the Fritzsch^\cite{f} matrix for both up
and down quarks.  This form is used for the up quarks only in
refs.^\cite{gj,hrr,ram,hh}}. \begin{eqnarray}
\left (
\begin{array}{clcr}
0 &a_5&0\\
a_5&0&b_5\\
0&b_5&c_5
\end{array}
\right )
\label{eq:f5}
\end{eqnarray}

\item
\begin{eqnarray}
\left (
\begin{array}{clcr}
0 &a_6&b_6\\
a_6&0&0\\
b_6&0&c_6
\end{array}
\right )
\label{eq:f6}
\end{eqnarray}

\end{enumerate}
\end{itemize}
where $a_i \lsim b_i \lsim c_i$ are constants determined by the
quark masses. Note that we have chosen the axes so that the
largest entry (approximately equal to the heaviest quark mass)
is in the (3,3) position. It may readily be verified that these
six forms are the complete set of possibilities up to relabeling
of the axes.

We will analyse all of these possibilities in turn for the up (or
the down) quark mass matrices. Then using eq(\ref{md}) we may
study the implications of three, two or one further zeroes in the
down (or the up) quark mass matrices, corresponding to a total
of 6, 5 or 4 texture zeroes.

The most predictive Ansatz has a total of 6 zeroes (3 in the up
and 3 in the down quark matrices) reducing the number of
parameters needed to specify the mass matrix. In terms of these
the 6 up and down quark masses and the CKM matrix elements must
be determined. In principle there are 6 measurable quantities in
the unitary CKM matrix but for small mixing angles this is
reduced to 4 leaving a total of 10 experimentally measurable
quantities. Mass matrices with texture are overconstrained
leading to the prediction of relations between these quantities
and it is our task to find which, if any, of these textures is
viable.

We are also able completely to analyse the case of five texture
zeroes, although one of the up or down quark rotations is not
completely determined. The analysis in this case proceeds exactly
as before because in this case too one of the up or down mass
matrices has three zeroes and is given by one of the set above.

The analysis of the case of four texture zeroes using the set of
matrices above is incomplete because four zeroes may also occur
when both the up and down matrices have just two zeroes. Below
we will discuss this case too, but we have not analysed its
implications fully because of the residual uncertainty in
determining both the up and down current quark basis.

\subsection{Parameterization of the mass matrices}
\label{sec:param}

As we will see, in the analytic analysis of possible textures it
is useful to parameterize the quark mass matrices in a way that
keeps track of the order of magnitude of the various components
of the mass matrices. This was done for the CKM matrix by
Wolfenstein^\cite{w}

\begin{equation}
{\bf V}_{CKM}=\left(\matrix{1-{\lambda2\over
2}&\lambda&A\lambda3 (\rho + i \eta)\cr -\lambda&1-
{\lambda2\over
2}&A\lambda2\cr
A\lambda3(1-\rho +i \eta)&-A\lambda2&1}\right)
\label{eq:wolf}
\end{equation}
where the small expansion parameter is $\lambda\approx .2$ the
(1,2) matrix element of the CKM matrix (approximately the Cabibbo
angle) and $ A\approx .9\pm .1$.
The assignment of the CP phase in eqn(\ref{eq:wolf})
is arbitrary, the only constraint is the invariance of $J{CP}$.
We now go one step further and parametrize the down quark masses
$\grave{a}$ {\it la} Wolfenstein. We choose\footnote{The
parameterisation of quark masses $\grave{a}$ {\it la} Wolfenstein
is not new, although here we take it to apply at the Unification
scale rather than at low energies. The first references are given
in^\cite{fish}.}

\begin{eqnarray}
{\bf Y}_{d}{Diag}=\left(\begin{array}{ccc}
m_d&0&0\\
0&\hat{m}_s/ \lambda2&0\\
0&0&\hat{m}_b/\lambda4
\end{array}
\right)
\label{eq:dd}
\end{eqnarray}
where
\begin{eqnarray}
\hat{m}_s\approx m_s \; \lambda2 \nonumber \\
\hat{m}_b\approx m_b \; \lambda4
\label{eq:sb}
\end{eqnarray}
are both of order $m_d$.
This  is a general parameterization of the down quark mass matrix
which is useful because it exhibits the order of magnitude of the
various elements. The analogous form for the up quarks is

\begin{equation}
{\bf Y}_{u}{Diag}=\left(\matrix{m_u&0&0\cr
0&\hat{m}_c/ \lambda4&0\cr
0&0&\hat{m}_t/\lambda8}\right)
\label{eq:ud}
\end{equation}
where
\begin{eqnarray}
\hat{m}_c\approx m_c \lambda4 \nonumber \\
\hat{m}_t\approx m_t \lambda8
\label{eq:ct}
\end{eqnarray}

\subsection{Evolution of the CKM matrix and the mass
eigenvalues.} \label{sec:C}
We have stressed that the experimentally observed quantities used
in our analysis are the CKM matrix and the eigenvalues of the
mass matrix and not the full mass matrices. For the analysis
presented above we would like to know these quantities at the
unification scale and thus we need to consider their radiative
corrections. In an elegant paper, Olechowski and
Pokorski^\cite{op} studied the renormalisation group
evolution of the CKM matrix and the masses and
we can use their results to get a rough idea of the size of the
radiative corrections.

Keeping the top and bottom Yukawa couplings only and neglecting
thresholds, the CKM matrix elements evolve as
\begin{eqnarray}
16\pi2\frac{d\mid V_{ij}\mid}{dt} & = & -
\frac{3c}{2}(h_t2+h_b2)\mid V_{ij} \mid, \; ij=13,31,23,32
\nonumber \\
\frac{d \mid V_{12} \mid}{dt} & \sim & {\cal O}(\lambda4)
\label{eq:rg1}
\end{eqnarray}
 Here $t=ln(Q/Q_0)$ where the elements are evaluated at the scale
Q, $h_t$ and $h_b$ are the Yukawa couplings and $c$ is a constant
determined by the couplings of the theory and is 2/3 in the MSSM.
The irreducible phase of the CKM matrix also evolves via the
equation
\begin{equation}
16\pi2\frac{d\mid J_{CP}\mid}{dt}=-3c(h_t2+h_b2)\mid
J_{CP}\mid
\label{eq:rg2}
\end{equation}
where we can choose $\mid J_{CP}\mid =
{\rm Im}(V_{23}V_{12}V_{13}{\star}V_{22}{\star})$.
If we substitute the Wolfenstein parameterisation,
eq(\ref{eq:wolf}) in eqs(\ref{eq:rg1}) and (\ref{eq:rg2}) we find
$\mid J_{CP} \mid=A2\lambda6(1-\lambda2/2)(\rho+i \eta)$ and
\begin{eqnarray}
16\pi2\frac{dA}{dt} & = & -\frac{3c}{2}(h_t2+h_b2)A \nonumber
\\
\frac{d\lambda}{dt} & \approx & 0\nonumber\\
\frac{d\rho}{dt} & = & 0\nonumber\\
\frac{d\eta}{dt} & = & 0\nonumber\\
\label{eq:da}
\end{eqnarray}

The beauty of this result is that only the $A$ parameter in
${\bf V}_{CKM}$ evolves on going from low to high scales making
the analysis including radiative corrections quite
straightforward. The remaining radiative corrections are to the
diagonal Yukawa couplings and have the form
\begin{equation}
16\pi2\frac{d(h_u/h_t)}{dt}  =  -\frac{3}{2}(b h_t2 +c h_b2)
(h_u/h_t)
\label{eq:rg3}
\end{equation}
and similarly for $(h_c/h_t)$.
\begin{equation}
16\pi2\frac{d(h_d/h_b)}{dt}  =  -\frac{3}{2}(c h_t2 +b h_b2)
(h_d/h_b)
\label{eq:rg4}
\end{equation}
and similarly for $(h_s/h_b)$. Here $b$=2 for the MSSM.

\subsubsection{Quantitative estimates.}

As we discuss in the next section, if we impose the GUT relation
$m_b=m_{\tau}$ at the GUT scale together with reasonable boundary
conditions on the SUSY breaking parameters, we need a relatively
large value of $h_t$ to get the required value of the running
mass $m_b = 4.25\pm0.1$ GeV evaluated at $m_b$. We can take $h_t$
to be approximately constant over the range $M_X$ down to $M_Z$.
It is then convenient to introduce the parameter $\chi$ defined
by $\chi=(M_X/M_Z){-h_t2/(16\pi2)}$. We have \begin{eqnarray}
\frac{A(M_X)}{A(M_z)} & = & \chi \nonumber \\
\frac{(h_d/h_b)(M_X)}{(h_d/h_b)(M_Z)} & = & \chi \nonumber \\
\frac{(h_u/h_t)(M_X)}{(h_u/h_t)(M_Z)} & = & \chi3
\label{eq:rga}
\end{eqnarray}
With the value of $h_t \approx 1.25$ used in our favourite
analysis of electroweak breaking (cf. Section 4.2) we have $\chi
\approx 0.7$. With this value the evolution of ${\bf V}_{CKM}$
up to $M_X$ requires A should be reduced by $\approx$30\%.
Relative to $h_d$ and $h_s$, $h_b$ should be increased by
$\approx$30\% at $M_X$ and relative to $h_u$ and $h_c$, $h_t$ is
increased by a factor 2.5$-$3 at $M_X$.

The beauty of this form is that the radiative corrections are
entirely specified in terms of the single parameter $\chi$.
Simply by varying $\chi$ it is easy to adjust  for other possible
values of the top quark coupling.
\section{Texture analysis}
\label{sec:tex}

We turn now to the results of applying the analytic analysis of
texture zeroes to the symmetric mass matrices. To illustrate the
method of Section 2 we consider the case where we start with a
specific texture structure with just {\it two} zeroes for the
down quark mass matrix.  This will illustrate the general method
capable of dealing with the 4 zero case and also, by setting one
of the elements zero, the  5 and 6 zero cases too.

\begin{eqnarray}
{\bf Y}_d=\left (
\begin{array}{clcr}
0 &a&0\\
a&b&c\\
0&c&d
\end{array}
\right )
\label{eq:deg}
\end{eqnarray}

Note that choosing c=0 gives the structure 2 (eq(\ref{eq:f2}))
and b=0 gives the structure 5  (eq(\ref{eq:f5})). In this
equation we have used five of the eight relative quark phases to
make the elements of ${\bf Y}_d$ all real. For $a \lsim b, c
\lsim d$ the rotation ${\bf R}_d$ is approximately given by

\begin{eqnarray}
{\bf R}_d \approx \left (
\begin{array}{clcr}
c_1 & -s_1 & 0 \\
s_1 & c_1 & 0 \\
0 & 0 & 1
\end{array}
\right ).
\left (
\begin{array}{clcr}
1 & 0 & 0 \\
0 & c_3 & -s_3 \\
0 & s_3 & c_3
\end{array}
\right )
\label{eq:vdeg}
\end{eqnarray}
where $c_{1,3}=\cos(\theta_{1,3})$, $s_{1,3}=\sin(\theta_{1,3})$
and
\begin{eqnarray}
s_1 \approx \frac{a d}{c2-b d}
\nonumber \\
s_3 \approx \frac{c}{d}
\label{eq:er}
\end{eqnarray}
and
\begin{eqnarray}
m_b & \approx & d \nonumber \\
m_{s} & \approx & -b +\frac{c{2}}{d}\nonumber \\
m_{d} & \approx & a{2}/m_{s}
\label{eq:md}
\end{eqnarray}

 Then applying the analogue of eq(\ref{eq:md}) for the up quark
mass matrix we have

\begin{eqnarray}
{\bf Y}_u & = & {\bf R}_{d}{\dagger}.P_d\;.\;{\bf
V}_{CKM}{\dagger}\;.\;P_u{\dagger}. {\bf
Y}_{u}{Diag}.P_u\;.\;{\bf V}_{CKM}\;.\;P_d{\dagger}.{\bf R}_{d}
\nonumber \\
        & = & P_d.P_d{\dagger}.{\bf
R}_{d}{\dagger}.P_d\;.\;{\bf
V}_{CKM}{\dagger}\;.\;P_u{\dagger}. {\bf
Y}_{u}{Diag}.P_u\;.\;{\bf V}_{CKM}\;.\;P_d{\dagger}.{\bf
R}_{d}.P_d.P_d{\dagger}
\label{eq:dprime}
\end{eqnarray}

It is convenient to compute ${\bf Y}_u$ in a different basis
${\bf Y}_u{\prime}=P_d{\dagger}.{\bf Y}_u.P_d$, and to absorb
the effect of $P_d{\dagger}.{\bf R}_{d}.P_d$ by allowing the off
diagonal elements  to be complex with phases $\phi$ and $\theta$.
It is now straightforward to determine the implications of zeroes
in the ${\bf Y}_u{\prime}$ matrix. To illustrate the method we
quote the result for ${\bf Y}_u{\prime}$ setting $s_3=0$ to keep
the algebra manageable

\begin{eqnarray}
{\bf Y}{\prime}_u(1,1) & = &  \frac{\hat{m}_c(1-
\epsilon)2+A2\hat{m}_t(\epsilon-(1-\rho+i\eta))2}{\lambda2}
+ \hat{m}_c\epsilon(1-\epsilon)+     m_u + {\rm O}(\lambda)
\nonumber \\
{\bf Y}{\prime}_u(1,2) & = & \frac{-\hat{m}_c(1-
\epsilon)+A2\hat{m}_t(\epsilon-(1-
\rho+i\eta))}{\lambda3} + {\rm O}(1) \nonumber \\
{\bf Y}{\prime}_u(1,3) & = &  \frac{- A\,\hat{m}_t
(\epsilon - (1 - \rho + i\,\eta ))}
        {\lambda5}  + {\rm
O}(\lambda){-1} \nonumber \\
{\bf Y}{\prime}_u(2,2) & = &
\frac{\hat{m}_c+A2\hat{m}_t}{\lambda4}+{\rm
O}(\lambda){-2} \nonumber \\
{\bf Y}{\prime}_u(2,3) & = & -\frac{A\,\hat{m}_t}{\lambda6} +
{\rm O}(\lambda){-4} \nonumber \\
{\bf Y}{\prime}_u(3,3) & = &  \frac{\hat{m}_t}{\lambda8} +
{\rm O}(1)
\label{eq:vue}
\end{eqnarray}
where, following the discussion of Section \ref{sec:param}, we
have parameterised the up quark mass matrix in terms of
$\hat{m}_c, \; \hat{m}_t$  and $\epsilon$ is  defined to be of
order one, $\epsilon\equiv s_1 e{i \phi} /\lambda$,
so that the expansion in $\lambda$ is well ordered.

It is now straightforward to use the form of eq(\ref{eq:vue}) to
find the implications of one or more texture zeroes in the up
quark mass matrix. From eq(\ref{eq:vue}) we see that it is not
possible for ${\bf Y}{\prime}_u(3,3)$ or ${\bf
Y}{\prime}_u(2,3)$  to be zero. The other
matrix elements may vanish for special values of the parameters.
For example if the (1,3) matrix element vanishes then, to leading
order in the Maclaurin series in $\lambda$

\begin{equation}
1-\rho +i \; \eta =\epsilon
\label{eq:v13}
\end{equation}

It is possible for the (1,1) matrix element to vanish
simultaneously with the (1,3) element. Inserting
eq(\ref{eq:v13}) in ${\bf Y}{\prime}_u(1,1)$ and working to the
next non-vanishing order in the expansion in powers of $\lambda$
gives \begin{equation}
\frac{\hat{m}_c(1-\epsilon)2}{\lambda2}+m_u=0 \label{eq:pp}
\end{equation}

At leading order this gives
\begin{equation}
\hat{m}_c(1-\epsilon)2=0
\label{eq:pp0}
\end{equation}
which, from eqs(\ref{eq:er}) and (\ref{eq:md}) gives the Gatto-
Sartori-Tonin-Oakes (GSTO)^\cite{gsto} relation
\begin{equation}
\lambda={ V}_{us}=\sqrt{m_d/m_s}
\label{eq:gsto}
\end{equation}

Solving eq(\ref{eq:pp}) to the next order gives
\begin{equation}
\mid { V}_{us} \mid = \lambda =(\frac{m_d}{m_s} +
\frac{m_u}{m_c}
+2 \sqrt{\frac{m_d \; m_u}{m_s \; m_c}} \cos \phi){\frac{1}{2}}
\label{eq:frit}
\end{equation}
where the phase $\phi$ comes from the phase of $\epsilon$.

Eq(\ref{eq:v13}) with eq(\ref{eq:pp})  gives
\begin{equation}
\rho2+\eta2=\lambda \; \mid\frac{m_u}{\hat{m}_c}\mid
=\lambda{-3} \mid \frac{m_u}{m_c} \mid
 \label{eq:re}
\end{equation}
Eqs(\ref{eq:frit}) and (\ref{eq:re}) are the consequences of two
texture zeroes in the (1,1) and (1,3) positions. From
eq(\ref{eq:frit}) we see that the expansion parameter,
$\lambda$, is of the correct order given by the GSTO relation but
with a small correction which determines the phase, $\phi$, of
$\epsilon$. From eq(\ref{eq:re}) we see that $\mid \rho + i \eta
\mid$ is predicted in terms of this expansion parameter to be
small, of order $\sqrt{\lambda}$. From eq(\ref{eq:v13}) and the
determination of $\phi$,  $\rho$ and $\eta$ may separately be
determined. This gives an example of a texture with 5 zeroes and
it may readily be verified that no further zeroes can be
obtained. We will return to a discussion of these results in
Section^\ref{sec:E}.

\subsection{Results of the analytic analysis}

The discussion presented in the last section illustrates the
method for studying 5 and 6 texture zeroes. Including $r\ne 0$
in eq(\ref{eq:dprime}) the most general case of 4 texture zeroes
can be also be analysed.  Using this method we have surveyed all
structures with 6 or 5 texture zeroes. A complete discussion of
these results will appear elsewhere but here we list just those
solutions that are consistent with present measurements of the
CKM matrix elements. Encouragingly there are solutions,
consistent with the hoped for simplicity in the mass matrices,
but the number of possible solutions is limited; no solutions
with 6 texture zeroes were found\footnote{In reference^\cite{dhr}
a solution with 6 texture zeroes was presented corresponding to
Solution 2 with $E{\prime}=0$. We do not include this in our
acceptable set of solutions as it leads to the prediction ${\bf
V}_{cb}=\sqrt{m_c/m_t}$ which is larger than present
indications that $m_t \le 180$GeV from LEP data^\cite{dallas}.
The non-zero entry in the
(2,3) position of the ${\bf Y}_d$ matrix of Solution 2 reduces
the value of ${\bf V}_{cb}$. However this is the nearest we get
to a 6 zero solution, setting one of elements to zero in any of
the other solutions leads to conflict with data. Note also that
our solution no.3 corresponds to the Ansatz of ref(\cite{giud}).}
and only five 5 texture zero solutions are consistent
with the measured masses and mixings. These are given in
Table^\ref{table:1}.

\begin{table}
\centering
\begin{tabular}{|c|c|c|} \hline
  &  &  \\
Solution & ${\bf Y}_u$ & ${\bf Y}_d$ \\
  &  &  \\
\hline
  &  &  \\
1 & $\left(
\begin{array}{ccc}
0 & C & 0 \\
C & B & 0 \\
0 & 0 & A
\end{array}
\right)$ & $\left(
\begin{array}{ccc}

0 & F & 0 \\
F{\star} & E & E{\prime} \\
0 & E{\prime} & D
\end{array}
\right)$
\\
 &  &  \\
\hline
 &  &  \\

2 & $\left(
\begin{array}{ccc}

0 & C & 0 \\
C & 0 & B \\
0 & B & A
\end{array}
\right)$ & $\left(
\begin{array}{ccc}

0 & F & 0 \\
F{\star} & E  & E{\prime} \\
0 & E{\prime \star} & D
\end{array}
\right)$
\\
  &  &  \\
\hline
  &  &  \\
3 & $\left(
\begin{array}{ccc}

0 & 0 & C \\
0 & B & 0 \\
C & 0 & A
\end{array}
\right)$ & $\left(
\begin{array}{ccc}

0 & F & 0 \\
F{\star} & E & E{\prime} \\
0 & E{\prime } & D
\end{array}
\right)$
\\
  &  &  \\
\hline
  &  &  \\
4 & $\left(
\begin{array}{ccc}

0 & C & 0 \\
C & B & B{\prime} \\
0 & B{\prime} & A
\end{array}
\right)$ & $\left(
\begin{array}{ccc}

0 & F & 0 \\
F{\star} & E & 0 \\
0 & 0 & D
\end{array}
\right)$
\\
  &  &  \\
\hline
  &  &  \\
5 & $\left(
\begin{array}{ccc}

0 & 0 & C \\
0 & B & B{\prime} \\
C & B{\prime} & A
\end{array}
\right)$ & $\left(
\begin{array}{ccc}

0 & F & 0 \\
F{\star} & E & 0 \\
0 & 0 & D
\end{array}
\right)$
\\
  &  &  \\
\hline
\end{tabular}
\caption{Symmetric textures. Approximate forms for the parameters
extracted from the various fits
are given in Table 2}
\label{table:1}
\end{table}

Note that the example presented above corresponds to Solution 4
with 5 texture zeroes (pairs of off-diagonal zeroes are counted
as one zero due to the symmetric structure assumed). After using
the freedom to choose the 9 independent quark phases there is one
phase left which we take to be the phase of the complex parameter
F. Thus we see this solution needs 7 real parameters and one
phase to describe the mass matrices and hence the 6 quark masses
and 4 CKM parameters. This implies two relations consistent with
our analysis showing $\rho$ and $\eta$ were determined. As a
bonus we found that the magnitude of the Cabibbo angle was
determined by the approximate GSTO relation.

The other solutions may be analysed in an equivalent way.
Solutions 1,3,4 and 5 have 8 parameters and hence yield two
relations. Solution 2 has an additional phase and hence only
gives one relation. The structure of these relations may be
algebraically complex (for example try solving the conditions
that the (1,1) and (1,2) matrix elements of eq(\ref{eq:vue})
simultaneously vanish - Solution 5). In order to check the
predictions it is convenient to make a best fit to the masses and
CKM matrix elements in terms of the non-zero elements. We will
do this in the next section in a more complete analysis including
the threshold effects in the renormalisation group analysis.

\section{Numerical Analysis of Yukawa Matrices}
\label{sec:D}

In the last section we discussed the analytic determination of
the implications of texture zeroes using the analytic form of the
radiative  corrections to
CKM matrix and the quark masses.  In this section we present the
results of a more complete  and exact
analysis where we study all possible 5 or 6 texture zero choices
including radiative corrections using a numerical solution to the
renormalisation group equations which correctly includes the
effects of thresholds in the analysis, omitted in our approximate
analytic treatment.   In this we begin at the scale $M_X$ with
a given pair of Yukawa matrices ${\bf Y}_u$ and ${\bf Y}_d$ and
evolve
down to low energies to examine the resulting fermion masses and
CKM matrix  elements.  Again we present the five solutions which
exhibit different  textures but are
all consistent with the data.

\subsection{Details of the Procedure}

The evolution of the ${\bf Y}_u$ mass matrix, for example, is
\begin{eqnarray}
16\pi2 \frac{d {\bf Y}_u}{dt} = \frac{3}{2}
(b{\bf Y}_u{\bf Y}_u{\dagger} +
c{\bf Y}_d{\bf Y}_d{\dagger}) {\bf Y}_u+
(\Sigma_u -A_u){\bf Y}_u
\label{eq:mevolv}
\end{eqnarray}
where $t=\ln Q$, $\Sigma_u =Tr [3{\bf Y}_u{\bf Y}_u{\dagger}
+3a{\bf Y}_d{\bf Y}_d{\dagger}+a{\bf Y}_e{\bf Y}_e{\dagger}]$,
$A_u=A_{u3} g2_3  +A_{u2}g2_2 +A_{u1} g2_1$.  Here ${\bf Y}_e$
is the lepton mass matrix and the values  coefficients $a,b,c,
A_{ui}$ depend on whether MSSM or SM is relevant.   Analogous to
eq(\ref{eq:mevolv})
there are equations for evolving the ${\bf Y}_{d,l}$ matrices.
Details may be found in
ref ^\cite{glt} for example.  In ref ^\cite{rr}  we derived
solutions for the SUSY spectrum  by
combining MSSM and unification of the gauge couplings and
demanding that the  electroweak symmetry is spontaneously broken
by radiative corrections.  We  continue
with this description here but ensuring that across each
individual threshold  of SUSY
particles, Higgs and quarks etc., not only do the gauge couplings
alter but  also the values
of the coefficients in eq(\ref{eq:mevolv}) and indeed individual
elements in ${\bf Y}_{u,d}$ and  ${\bf Y}_e$.

In this procedure the parameters at the unification scale
determining the supersymmetric mass spectrum are the soft SUSY
breaking parameters $m_{1/2}, m_0,  \mu_0, A, B$. The allowed
ranges of these parameters may be constrained by the need to
obtain the correct scale of electroweak breaking without fine
tuning, consistency with dark matter abundance and unification
of gauge couplings. In our estimates of threshold effects we will
use values consistent with these constraints.

We choose forms for ${\bf Y}_{u,d}$ at $M_X$, either one of the
six three-zero  types given
by eqns (5)-(10) or one of the 12 possible two-zero types derived
from them.  Provided ${\bf Y}_e$ is small it plays a negligible
role in the evolution equation \ref{eq:mevolv} and can be ignored
in an analysis of the quark structure. However, as has been
stressed in ^\cite{dhr}, the simple Georgi-Jarlskog Ansatz is
very successful in describing the lepton mass matrix and so, for
completeness, we analyse this structure too. Motivated by this
Ansatz the form  for ${\bf Y}_e$ we take is
governed by the choice for ${\bf Y}_d$, taking the same zero
texture but multiplying the (2,2)
element by the factor 3 in order to give  the relations for  the
eigenvalues at the scale $Q=M_X$ : $ h_\tau=h_b, h_\mu=3h_s,
h_e=h_d/3$\footnote {We have chosen to include a discussion of
the lepton masses to illustrate that all the quark mass
structures of interest here may be combined with reasonable
lepton masses, given the simple G-J Ansatz. However, the analysis
of of quark mass matrices is essentially independent of this
Ansatz and does not change if we choose not to include lepton
masses in the analysis.}

Given the initial conditions we now evolve down to low energies
via the renormalisation group equations all  quantities including
the matrix elements of ${\bf Y}_{u,d}$ and ${\bf Y}_e$ and
attempt to fit  the values
of $m_e,m_\mu,m_\tau,m_d,m_s,m_b,m_u,m_c$ and the CKM matrix
elements. We have included the analysis of lepton masses for
completeness but we stress once more that our assumption of a
Georgi-Jarlskog-like Ansatz in the lepton sector plays a
negligible role in determining the quark structure.

\subsection{Results of analysis}

As discussed above, we actually include the lepton masses in
generating our solutions. In particular we assume $h_b/h_\tau =1$
at $Q=M_X$ and to achieve a value for this ratio at low energy
around 2.4 we find we need $h_t$ to be relatively large.
Depending on our particular choice for the parameters of the MSSM
at $Q=M_X$  ($m_{1/2}$, $m_0$, etc.) the values of $h_t$ are in
the range 1$-$1.5 and this value enters into all the relevant
evolution equations of the masses. Predictions for $\tan \beta$
depend in particular on the assumed values for $A$, $B$ at
$Q=M_X$ and this uncertainty (together with the uncertainty on
$m_b/m_\tau$) translates into  a range of values for $m_t$ of
145$-$185 GeV. The masses of the other fermions are independent
of the precise value of $m_t$ however.

Altogether we find five solutions for the Yukawa matrices at
$Q=M_X$ which are  consistent with the low energy fermion masses
and CKM matrix elements.  The structure of the five
forms are listed in table^\ref{table:1}.  Rather than list the
numeric values we present,  in table 2,
approximations to the matrix elements in powers of
$\lambda(\simeq$ 0.22),  the small
parameter in eqn (11).  In table 3 we list the results of the
five solutions  for the fermion
masses and CKM matrix elements.

Note that in our five solutions there is a single candidate
structure, for  ${\bf Y}_d$, namely the
form of eq(\ref{eq:deg}), which for certain cases reduces to the
form of  eq(\ref{eq:f2}).
For ${\bf Y}_u$ the five
solutions correspond to the forms
eqs(\ref{eq:f2},^\ref{eq:f4},^\ref{eq:f5},^\ref{eq:f6},  (6),
(8), (9), (19) and one related
to eq(\ref{eq:deg}) by an
interchange of axes.

\begin{table}
\centering
\begin{tabular}{|c|c|c|} \hline
  &  &  \\
Solution & ${\bf Y}_u$ & ${\bf Y}_d$ \\
  &  &  \\
\hline
  &  &  \\
1 & $\left(
\begin{array}{ccc}

0 & \sqrt{2}\lambda6 & 0 \\
\sqrt{2}\lambda6 & \lambda4 & 0 \\
0 & 0 & 1
\end{array}
\right)$ & $\left(
\begin{array}{ccc}

0 & 2\lambda4 & 0 \\
2\lambda4 & 2\lambda3 & 4\lambda3 \\
0 & 4\lambda3 & 1
\end{array}
\right)$
\\
  &  &  \\
\hline
  &  &  \\
2 & $\left(
\begin{array}{ccc}

0 & \lambda6 & 0 \\
\lambda6 & 0 & \lambda2 \\
0 & \lambda2 & 1
\end{array}
\right)$ & $\left(
\begin{array}{ccc}

0 & 2\lambda4 & 0 \\
2\lambda4 & 2\lambda3 & 2\lambda3 \\
0 & 2\lambda3 & 1
\end{array}
\right)$
\\
  &  &  \\
\hline
  &  &  \\
3 & $\left(
\begin{array}{ccc}

0 & 0 & \sqrt{2}\lambda4 \\
0 & \lambda4 & 0 \\
\sqrt{2}\lambda4 & 0 & 1
\end{array}
\right)$ & $\left(
\begin{array}{ccc}

0 & 2\lambda4 & 0 \\
2\lambda4 & 2\lambda3 & 4\lambda3 \\
0 & 4\lambda3 & 1
\end{array}
\right)$
\\
  &  &  \\
\hline
  &  &  \\
4 & $\left(
\begin{array}{ccc}

0 & \sqrt{2}\lambda6 & 0 \\
\sqrt{2}\lambda6 & \sqrt{3}\lambda4 & \lambda2 \\
0 & \lambda2 & 1
\end{array}
\right)$ & $\left(
\begin{array}{ccc}

0 & 2\lambda4 & 0 \\
2\lambda4 & 2\lambda3 & 0 \\
0 & 0 & 1
\end{array}
\right)$
\\
  &  &  \\
\hline
  &  &  \\
5 & $\left(
\begin{array}{ccc}

0 & 0 & \lambda4 \\
0 & \sqrt{2}\lambda4 & \frac{\lambda2}{\sqrt{2}} \\
\lambda4 & \frac{\lambda2}{\sqrt{2}} & 1
\end{array}
\right)$ & $\left(
\begin{array}{ccc}

0 & 2\lambda4 & 0 \\
2\lambda4 & 2\lambda3 & 0 \\
0 & 0 & 1
\end{array}
\right)$
\\
  &  &  \\
\hline
\end{tabular}
\caption{Approximate forms for the symmetric textures using the
parameterisation of Section 2. }
\label{table:2}
\end{table}

\begin{table}
\begin{center}
\begin{tabular}{|c|cccccc|}\hline
  & & & & & & \\
Solution  & 1 & 2 & 3 & 4 & 5 & Experiment^\cite{leut,buras} \\
  & & & & & & \\
\hline
  & & & & & & \\
$m_d$ & .0075 & .0072  & .0082  & .0071 & .0071 & .0055 $-$ .0115\\
$m_s$ & 0.159 & 0.170 & 0.164 & 0.169 & 0.169 & 0.105 $-$ 0.230\\
$m_b$ & 4.23 &  4.23 & 4.23 & 4.19 & 4.20 & 4.15 $-$ 4.35\\
$m_u$ & .0042 & .0040  & .0045  & .0044 & .0040 & .0031 $-$ .0064\\
$m_c$ & 1.27 & 1.28 & 1.27 & 1.30 & 1.20 & 1.22 $-$ 1.32\\
  & & & & & & \\
$|V_{ud}|$ & .9756 & .9754 & .9759 & .9756 & .9758 & .9747 $-$ .9759\\
$|V_{us}|$ & .2197 & .2203 & .2180 & .2197 & .2185 & .218 $-$ .224\\
$|V_{ub}|$ & .0029 & .0029 & .0034 & .0029 & .0040 & .003 $-$ .008\\
$|V_{cd}|$ & .2195 & .2201 & .2178 & .2195 & .2184 & .218 $-$ .224\\
$|V_{cs}|$ & .9744 & .9743 & .9747 & .9744 & .9748 & .9734 $-$ .9752\\
$|V_{cb}|$ & .0483 & .0471 & .0500 & .0490 & .0448 & .035 $-$ .047\\
$|V_{td}|$ & .0100 & .0096 & .0110 & .0098 & .0078 & .006 $-$ .018\\
$|V_{ts}|$ & .0474 & .0462 & .0488 & .0481 & .0443 & .035 $-$ .047\\
$|V_{tb}|$ & .9988 & .9989 & .9987 & .9988 & .9990 & .9987 $-$ .9994\\
  & & & & & & \\
$\phi_{CP}$ & $1110$ & $650$ & $960$ & $1170$ & $490$ & $250 - 1600$\\
  & & & & & & \\
\hline
\end{tabular}
\caption{Resulting values for masses and CKM matrix elements for
the five  solutions of table 1.  Masses in GeV. For the solutions
considered in this analysis, $m_t \sim $ 180 GeV but this value
could easily be decreased to around 150 GeV as a result of the
uncertainty in the value of $\tan \beta$ and $m_b$.}
\end{center}
\label{table:numbers}
\end{table}

\section{Discussion of the supersymmetric textures.}
\label{sec:E}

\subsection{Structure of the ${\bf Y}_d$ matrix at $M_X$.} All
of the textures found in the last section have the form
\begin{equation}
{\bf Y}_d=m_b\left( \begin{array}{ccc}
0 & \alpha \; \lambda4   & 0 \\
\alpha \; \lambda4 & \beta \; \lambda3 & \gamma \; \lambda3
\\
0 & \gamma \;  \lambda3 & 1
\end{array}
\right)
\label{eq:gdf1}
\end{equation}
where the parameters $\alpha$, $\beta$, and $\gamma$ are of
$O(1)$
\begin{eqnarray}
\alpha & = & \frac{\sqrt{\hat{m}_sm_d}}{\hat{m}_b}
\nonumber \\
\beta & = & \frac{\hat{m}_s}{\hat{m}_b}
\label{eq:cf1}
\end{eqnarray}
 The results of the last section determine the constants $\alpha,
\; \beta$ and $\gamma$ : $\alpha=\beta=2$, $\gamma=2N$
for $N=0,1,2$. To make a connection with the analysis of Section
3 we note that

\begin{eqnarray}
\frac{\hat{m}_s}{\hat{m}_b} & = & 2 \nonumber \\
\frac{\hat{m}_s}{m_d} & = & 1 \nonumber \\
s_3 & = & \frac{1}{2} N \lambda2 \nonumber \\
s_1 & = & \lambda
\label{eq:r1}
\end{eqnarray}
where $s_1$ and $s_3$ are the mixing angles of
eq(\ref{eq:vdeg}), the general structure of ${\bf R}_d$ given in
this equation being relevant because the ${\bf Y}_d$ matrix of
Table^\ref{table:1} is always of the form of
eq(\ref{eq:deg}).

\subsection{Structure of the ${\bf Y}_u$ matrix at $M_X$.}
Similarly the solutions 1,2,and 4 for the ${\bf Y}_u$ matrix have
the form \begin{equation}
{\bf Y}_u=m_t\left( \begin{array}{ccc}
0 & \alpha{\prime} \; \lambda6 & 0 \\
\alpha{\prime} \; \lambda6 & \beta{\prime} \; \lambda4 &
\gamma{\prime} \; \lambda2 \\
0 & \gamma{\prime} \;  \lambda2 & 1
\end{array}
\right)
\label{eq:guf2}
\end{equation}
where
\begin{eqnarray}
\alpha{\prime} & = & \frac{\sqrt{m_c{\prime}m_u}}{m_t{\prime}}
\nonumber \\
\beta{\prime} & = & \frac{m_c{\prime}}{m_t{\prime}}
\label{eq:cf2}
\end{eqnarray}
and we have chosen the (2,3) matrix element in eq(\ref{eq:cf2})
as is found in our fits. The results of the last section
determine the constants $\alpha{\prime}, \; \beta{\prime}$ and
$\gamma{\prime}$

With the phases chosen so that the ${\bf Y}_u$ matrix is real the
matrix needed to diagonalise it has the form

\begin{equation}
{\bf R}_u=\left(\begin{array}{ccc}
c_2 & s_2 & 0 \\
-s_2 & c_2 & 0 \\
0 & 0 & 1
\end{array}
\right).
\left(\begin{array}{ccc}
1 & 0 & 0 \\
0 & c_3 & s_3 \\
0 & -s_3 & c_3
\end{array}
\right)
\label{eq:vuf}
\end{equation}

Using ${\bf R}_d$ given in eq(\ref{eq:vdeg})  we may determine
the CKM matrix in the basis in which ${\bf Y}_d$ is diagonal, by
\begin{equation}
{\bf V}_{CKM}={\bf R}_u \; P_u \; {\bf R}_d{\dagger}
\label{eq:ckmf}
\end{equation}
where $Pu$ is the diagonal matrix of phases which relates the
basis in which ${\bf Y}_{u}$ is diagonal to the basis in which
${\bf Y}_{d}$ is real. Just how many phases are required to
determine $Pu$ depends on the number of texture zeroes. An
overall phase is irrelevant so we may write

\begin{equation}
P_u=\left(\begin{array}{ccc}
e{i \phi} & 0 & 0 \\
0 & e{i \phi} & 0 \\
0 & 0 & 1
\end{array}
\right).
\left(\begin{array}{ccc}
1 & 0 & 0 \\
0 & e{i \theta} & 0 \\
0 & 0 & e{i \theta}
\end{array}
\right)
\label{eq:pu}
\end{equation}

Thus there are at most two phases ($\phi$ and $\theta$) giving
for ${\bf V}_{CKM}$

\begin{eqnarray}
{\bf V}_{CKM} \approx \left(\begin{array}{ccc}
c_1c_2-s_1s_2e{-i\phi}  &  s_1e{i\phi}+
c_1 s_2  &  s_2(s_3-s_4e{i\theta})  \\
 -c_1s_2-s_1e{-i\phi}  &
-s_1s_2e{i\phi}+c_1c_2c_3c_4+s_3s_4e{-i\theta}  &
s_3-s_4e{i\theta}  \\
 s_1e{-i\phi}(s_3-s_4e{-i\theta})  &
-c_1(s_3-s_4e{i\theta})  &
c_3c_4 + s_3s_4e{i\theta}
\end{array}\right)
\label{eq:ckmff}
\end{eqnarray}
giving
\begin{eqnarray}
\mid { V}_{us} \mid & = & (s_12+s_22+2 s_1 s_2
\cos\phi){\frac{1}{2}} \nonumber \\
\mid { V}_{cb} \mid & = & (s_32+s_42+2 s_3 s_4
\cos\theta){\frac{1}{2}} \nonumber \\
\frac{|{ V}_{ub}|}{|{ V}_{cb}|} & = & s_2
\label{eq:me}
\end{eqnarray}

To illustrate the structure we derive for the ${\bf V}_{CKM}$
matrix that results from one of the texture structures of
Section^\ref{sec:tex} we consider solution 4. In this case  there
are two phases left after field redefinition. The up mixing
angles are $s_3=0$ and $s_2=\sqrt{\frac{m_u}{m_c}}$ and (cf.
eq(\ref{eq:ckmff})) the vanishing of $s_3$ means that one of the
phases is irrelevant. From eqs(\ref{eq:er}) and (\ref{eq:me}) we
may then derive eq(\ref{eq:frit}) as must be the case since ${\bf
Y}_{u}$ has a zero in the (1,1) position. We also have from
eq(\ref{eq:me}) the result
$\frac{{\bf V}_{ub}}{{\bf V}_{cb}}=\sqrt{\frac{m_u}{m_c}}$. This
result also follows from eq(\ref{eq:re})
again as is expected since ${\bf Y}_{u}$ has a zero in
the (1,3) position.

\subsubsection{Comparison with analytic analysis.}

Here we compare the full results of Section^\ref{sec:D} with the
expectation following from the analytic solution of the
renormalisation group equations given in Section^\ref{sec:C}.
\begin{table}
\centering
\begin{tabular}{|c|c|c|c|c|} \hline
  & & & &  \\
 Solution & $ m_c/m_t $&$ m_u/m_c  $&$ V_{cb} $&$ V_{ub}/V_{cb}$\\
 & & & & \\ \hline
 & & & & \\
$1 $&$ \frac{\lambda4}{\chi3} $&$ 2 \; \lambda4 $&$ \frac{4
\; \lambda3}{\chi} $&$ \sqrt{2} \; \lambda2$ \\
 & & & & \\
$2 $&$ \frac{\lambda4}{\chi3} $&$ \lambda4 $&$
\frac{\lambda2(1-2 \; \lambda)}{\chi} $&$ \lambda2$ \\
 & & & & \\
$3 $&$ \frac{\lambda4}{\chi3} $&$ 2 \; \lambda4 $&$ \frac{4
\; \lambda3}{\chi} $&$ \frac{1}{2\sqrt{2}} \; \lambda$ \\
 & & & & \\
$4 $&$ \frac{(\sqrt{3}-1)\lambda4}{\chi} $&$
(\sqrt{3}+2)\lambda4 $&$ \frac{\lambda2}{\chi3} $&$
{\frac{\sqrt{2}}{\sqrt{3}-1}} \; \lambda2 $\\
 & & & & \\
$ 5 $ & $\frac{(\sqrt{2}-\frac{1}{2}) \;\lambda4}{\chi3}$ & $
\frac{\sqrt{2}}{(\sqrt{2}-\frac{1}{2})2}\lambda4 $ &
$\frac{\frac{1}{\sqrt{2}}\lambda2}{\chi} $ & $ \sqrt{2} \;
\lambda2 $\\
 & & & & \\
\hline
\end{tabular}
\caption{Results following from the five symmetric texture
solutions.}
\label{table:3}
\end{table}

\begin{table}
\centering
\begin{tabular}{|c|c|c|c|c|} \hline
 & & & & \\
 Solution & $ m_c/m_t $&$ m_u/m_c  $&$ V_{cb} $&$ V_{ub}/V_{cb}$\\
 & & & & \\
\hline
1 & 0.0067 & 0.0046 & 0.060 & 0.068 \\ \hline
2 & 0.0067 & 0.0023 & 0.038 & 0.048 \\ \hline
3 & 0.0067 & 0.0046 & 0.060 & 0.078 \\ \hline
4 & 0.0049 & 0.0087 & 0.068 & 0.040 \\ \hline
5 & 0.0061 & 0.0030 & 0.048 & 0.068 \\ \hline
\end{tabular}
\caption{Results following from the five symmetric texture
solutions using $\lambda$=0.22.}
\label{table:3n}
\end{table}

For all five solutions we have, for the parameters evaluated at
a scale $M_Z$
\begin{eqnarray}
V_{us}\approx V_{cd} & \approx & \lambda \nonumber \\
\frac{m_s}{m_b} & = & \frac{2\; \lambda3}{\chi} \nonumber \\
\frac{m_d}{m_s} & = & \lambda2
\label{eq:sol1}
\end{eqnarray}
where $\chi=(M_X/M_Z){-h_t2/(16\pi2)}$ and we have taken $h_t$
approximately constant as in eq(\ref{eq:rga}).
The results for the remaining elements of ${\bf V}_{CKM}$ are
given in Table^\ref{table:3}.

These are the values evaluated at the scale $M_Z$. For comparison
we quote the experimental results^\cite{buras}
\begin{eqnarray}
\frac{m_s}{m_b} & = & 0.03-0.07 \nonumber \\
\frac{m_d}{m_s} & = & 0.04 - 0.067 \nonumber \\
V_{cb} & = & 0.025 - 0.050 \nonumber \\
\frac{V_{ub}}{V_{cb}} & = & 0.05-0.13 \nonumber \\
\frac{m_c}{m_t} & \approx &  0.0072 \; ({\rm for} \;
m_t=180\;{\rm GeV}) \nonumber \\
\frac{m_u}{m_c} & = & 0.003-0.005
\label{eq:experiment}
\end{eqnarray}

This is to be compared with the texture results obtained using
the best value for $\lambda=0.22$. All solutions give
$V_{us}$=0.22 and $\frac{m_d}{m_s}$=0.05 and
$\frac{m_s}{m_b}$=0.03, in agreement with the experimental
results.

  For the other quantities we have the results given in
Table^\ref{table:3n}.

It may be seen all quantities are reasonably close to
the experimental values, given the simple analytic form,  which
ignores threshold  effects, taken for
the evolution of the CKM matrix elements and Yukawa couplings.
Thus we see the analytic approximation gives quite a reliable
guide to the results of the complete numerical analysis for all
quantities.

\section{Discussion and Conclusions}
\label{sec:F}
We have determined all possible forms of symmetric quark mass
matrices having a total of five or six texture zeroes which are
consistent with the measured values of the quark masses and
mixing angles. It is encouraging for the idea of unification that
there are such solutions corresponding to simplicity at the
unification scale. This simplicity extends to the lepton quark
matrices for the lepton masses are consistent with the Georgi
Jarlskog relations at the unification scale.

With the present measurements of quark masses and mixing angles,
we find several candidate solutions, corresponding to the
discrete ambiguity in determining the current quark basis. In
detail these solutions give different predictions relating the
masses and mixing angles and so may be distinguished by improved
measurements of the CKM matrix elements. Perhaps the most useful
outcome of this analysis is the determination, cf. Table 3, of
the level of accuracy needed for the discrimination between
solutions.

The obvious question raised by these textures is what underlying
theory can lead to such structure? Although a detailed answer to
this question lies beyond the scope of this paper we cannot
resist drawing some conclusions from the general structure found
in the various solutions. The parameterisation of the quark (and
lepton) mass matrices that was suggested by the hierarchy of
masses and mixing angles is (cf eqs(\ref{eq:wolf}),(\ref{eq:dd}),
and (\ref{eq:ud})) a perturbative expansion in $\lambda$. This
structure strongly suggests to us an underlying symmetry broken
by terms of $O(\lambda )$. In the limit the symmetry is exact
only the third generation is massive and all mixing angles are
zero. Symmetry breaking terms gradually fill in the mass matrices
generating an hierarchy of mass scales and mixing angles. Of
course the idea of such a symmetry structure is not
new^\cite{demo} and essentially all attempts to provide an
explanation of the quark and lepton masses rely on broken
symmetry, although they may not emphasise its role. However the
realisation of the importance of an underlying symmetry leads to
a discussion of the general properties such a solution must have
and indeed these properties seem to be quite promising in
explaining the form of the
structures found.

To illustrate this let us first discuss the two generation case.
We consider the simplest possible symmetry based on an $U(1)$
symmetry or on a $Z_N$ discrete subgroup. (Such symmetries are
common in Grand Unification or compactified string theories. Of
course the structure that emerges may also be derived from larger
symmetries.) With our assumption that the mass matrices are
symmetric we must take the left- and the right- handed components
to transform in the same way under this symmetry. Moreover the
$SU(2)$ symmetry requires that the up and down quarks must have
the same transformation properties. Thus, without any loss of
generality we may take the transformation properties of the
quarks to be
\begin{eqnarray*}
c_{L,R},\; s_{L,R} \rightarrow c_{L,R},\; s_{L,R} \\
u_{L,R},\; d_{L,R} \rightarrow  \alpha \; u_{L,R},\; \alpha \;
d_{L,R}
\label{eq:tr1}
\end{eqnarray*}
where $\alpha = exp(i2\pi/N)$.

The transformation properties of the elements of the quark mass
matrix then have the form
\begin{equation}\left (
\begin{array}{cc}
u_L & c_L
\end{array}
\right )
{}.
\left (
\begin{array}{cc}
\alpha2 & \alpha \\
\alpha & 1
\end{array}
\right )
{}.
\left (
\begin{array}{c}
u_R \\
c_R
\end{array}
\right )
\; \; \;
\left (
\begin{array}{cc}
d_L & s_L
\end{array}
\right )
{}.
\left (
\begin{array}{cc}
\alpha2 & \alpha \\
\alpha & 1
\end{array}
\right )
{}.
\left (
\begin{array}{c}
d_R \\
s_R
\end{array}
\right )
\label{eq:tr2}
\end{equation}

The form of the mass matrices depends on the transformation
properties of the Higgs, $H_{1,2}$, which couple to the up and
down quarks respectively and generate their masses once the
electroweak symmetry is broken. If they are singlets under the
$Z_N$ group then, from eq(\ref{eq:tr2}), we see that only the c
and s quarks acquire mass. If, however, the $Z_N$ symmetry is
broken by the vacuum expectation value, $x$,  of a field $\Theta$
transforming as $\Theta \rightarrow \bar{\alpha} \Theta $ then
we may expect corrections to the mass matrix to occur due to
higher dimension terms coupling the quarks to the combination of
fields $H_{1,2} \Thetan$ for some integer n, giving a mass
matrix of the form
\begin{equation}
{\bf Y}=\left (
\begin{array}{cc}
\frac{x2}{M_12} & \frac{x}{M_2} \\
\frac{x}{M_2} & 1
\end{array}
\right )
\label{eq:mm1}
\end{equation}
where $M_{1,2}$ are the masses associated with the scale of new
physics generating the higher dimensional terms.

Another way that such an hierarchy may arise is through the
mixing of $H_{1,2}$ with other Higgs states $H_{1,2}{a,b}$
transforming as $\bar{\alpha}$ and $\bar{\alpha}2$. Then the
light Higgs state will be a mixture $H_{1,2}{Light}\approx
H_{1,2}+\frac{x}{M_2}H_{1,2}{a}+\frac{x2}{M_12}H_{1,2}{b}$,
where $x$ is now the vev of a field $\bar{\Theta}$ transforming as
$\bar{\Theta} \rightarrow \alpha \bar{\Theta} $ and  $M_{1,2}$
are the masses of intermediate states mixing $H_{1,2}$ with the
other Higgs states. This again gives the structure of
eq(\ref{eq:mm1}).

It is now easy to see how all of the structures of Table 2 may
arise. Assuming, for simplicity, a single scale of new physics,
$M_{1,2}=M$, then
\begin{equation}
{\bf Y}=
\left(
\begin{array}{cc}
\lambda2 &\lambda \\
\lambda & 1
\end{array}
\right )
\; \; \;
\lambda= \frac{x}{M}
\label{eq:mm2}
\end{equation}
If there should be no additional field $H_{1,2}b$ transforming
as
$\bar{\alpha}2$, then
\begin{equation}
{\bf Y}=
\left(
\begin{array}{cc}
0 &\lambda \\
\lambda & 1
\end{array}
\right )
\label{eq:mm3}
\end{equation}
If instead there is no additional field $H_{1,2}a$ transforming
as $\bar{\alpha}$ then
\begin{equation}
{\bf Y}=
\left(
\begin{array}{cc}
\lambda2 &0 \\
0 & 1
\end{array}
\right)
\label{eq:mm4}
\end{equation}
The remaining structure encountered in Table 2 has the form
\begin{equation}
{\bf Y}=
\left(
\begin{array}{cc}
\lambda & \lambda \\
\lambda & 1
\end{array}
\right)
\label{eq:mm5}
\end{equation}
and it too may also be generated. For example if the discrete
symmetry is $Z_3$ then $\alpha2=\bar{\alpha}$ in
eq(\ref{eq:tr2}). Then the structure of eq(\ref{eq:mm5}) results
if the vev of $\Theta$ develops along a ``D-flat'' direction
$<\Theta>=<\bar{\Theta}>=x$, where $\bar{\Theta}$ is a field in
the conjugate representation to $\Theta$\footnote{The spontaneous
breaking of symmetries at high scales in supersymmetric theories
must proceed along such flat directions.}. Another possibility
discussed below is that the form of eq(\ref{eq:mm5}) comes from
eq(\ref{eq:mm1}) because the masses $M_i$ are not degenerate.

Thus we may easily generate any of the substructures found in
Table 2. Indeed such structures are the natural expectations in
any underlying theory, such as a Grand Unified theory or a
compactified string theory, which possesses additional symmetries
of the type discussed.

It is straightforward to extend this discussion to the three
generation case. Rather than present a general analysis let us
just illustrate the possibilities by presenting a symmetry
structure which leads to one of the solutions found in Table 2.
We assume that the underlying theory yields a three family
spectrum of quarks with transformation properties
\begin{eqnarray*}
t_{L,R},\; b_{L,R} \rightarrow t_{L,R},\; b_{L,R} \\
c_{L,R},\; s_{L,R} \rightarrow \alpha c_{L,R},\; \alpha s_{L,R}
\\
u_{L,R},\; d_{L,R} \rightarrow  \bar{\alpha}4 \; u_{L,R},\;
\bar{\alpha}4 \; d_{L,R}
\label{eq:tr3}
\end{eqnarray*}
We further assume that, in addition to the light $Z_N$ singlet
Higgs field, $H_1$, needed for electroweak symmetry breaking,
there are massive Higgs fields $H_1{a,b}$ transforming under
$Z_N$ as $\bar{\alpha}$ and $\alpha3$ with which the singlet
field may mix. Assuming only a single scale of new physics the
resulting light Higgs has the form $H_1{Light}\approx
H_1+\bar{\theta}H_1a/M+\theta3 H_1b/M3$ giving for the up
quark mass matrix the form (with $\lambda2=x/M$)
\begin{equation}
{\bf Y}_u\approx \left (
\begin{array}{ccc}
0 & \lambda{6} & 0\\
\lambda6 & 0 & \lambda{2} \\
0 & \lambda2 &1
\end{array}
\right )
\label{eq:mm6}
\end{equation}
which is of the form of the up quark mass matrix of
case 2 in Table 2. (In
deriving this result we have assumed that the vev of $\Theta$
develops along a ``D-flat'' direction
$<\Theta>=<\bar{\Theta}>=x$.)

What this example shows is how the pattern of fermion masses and
mixing angles, as encoded in the mass matrices, may simply be
related to the multiplet structure of the underlying theory. As
we have seen the apparently complicated pattern may result from
a relatively simple multiplet structure. To complete this example
we need to explain why the down quarks have a different mass
matrix. Again this proves to be relatively simple to explain on
the basis of the multiplet structure. Any difference between the
up and down quark mass matrices comes from different mixings of
the Higgs, $H_{1,2}$, giving masses to the up and down quarks
respectively. For example we suppose that in the $H_2$ sector
there is an additional massive Higgs field $H_2c$
transforming as $\alpha2$. We further assume that the massive
Higgs fields transforming as $\bar{\alpha}3$ and $\alpha2$
receive {\it their} mass only at order $x$ via the $\Theta$
vev\footnote{This will happen naturally if the fields with which
they couple to obtain their mass have the appropriate
transformation properties under the $Z_N$ group.}, so that the
mixing of these fields to the light singlet $H_2$ will be
enhanced by a factor $M/x$. The resulting light Higgs field is
$H_2{Light}\approx H_2+\bar{\theta}H_2a/M+\theta3 H_2b/xM
+\theta2 H_2c/xM2$ giving a down quark mass matrix of the form
\begin{equation}
{\bf Y}_d\approx \left (
\begin{array}{ccc}
0 & \lambda{4} & 0\\
\lambda{4} & \lambda{2} & \lambda{2} \\
0 & \lambda2 &1
\end{array}
\right )
\label{eq:mm7}
\end{equation}
Up to corrections of $O(1)$, which may be expected in generating
the higher dimension terms, this is of the form of the down quark
mass matrix in Table 2. Thus we have constructed an example
which, by assuming a definite multiplet and symmetry structure,
generates the full texture structure of case 2 of Table 2 in
terms of a single expansion parameter, $x$.

Of course the next step is to identify the underlying theory
which leads to the appropriate multiplet and symmetry structure.
It is known in compactified string theories that definite family
structures for quarks and leptons may emerge and that they
possess definite transformation properties under discrete
symmetries of the type discussed above. Indeed specific models
have been analyzed which do lead to structures in the fermion
mass matrices of the type just discussed^\cite{ggr}. Similarly
additional broken gauge symmetries may lead to the type of mass
structure discussed above^\cite{dhr2}.

In conclusion, the fact that simple textures for the quark and
lepton masses can describe in detail the quark and lepton masses
and mixings lends further circumstantial evidence in favour of
an underlying unified theory. It is to be hoped that improvements
in experimental measurements of the CKM matrix elements will
further refine this evidence. To us the resultant structure
strongly suggests a (broken) symmetry explanation of the
structure of the type which naturally arises in GUTs or
compactified string unification and encourages us in the search
for a definite theory.

\newpage

\end{document}